\documentclass[aps,10pt,pra,superscriptaddress,showpacs,floatfix,twocolumn,longbibliography,nofootinbib]{revtex4-1}
\usepackage{graphicx}
\usepackage{dcolumn}
\usepackage{bm}
\usepackage{physics}
\usepackage[toc, page]{appendix}
\usepackage{epstopdf}
\usepackage{comment}
\usepackage{amssymb}
\usepackage{hyperref}
\usepackage{subfigure}

\hypersetup{
    colorlinks=true,       
    linkcolor=cyan,          
    citecolor=magenta,        
    filecolor=magenta,      
    urlcolor=cyan,           
    runcolor=cyan
}
\usepackage[pdftex]{color}

\newcommand{\beq}{\begin{equation}}
\newcommand{\eeq}{\end{equation}}
\newcommand{\beqnn}{\begin{equation*}}
\newcommand{\eeqnn}{\end{equation*}}
\newcommand{\bea}{\begin{eqnarray}}
\newcommand{\eea}{\end{eqnarray}}
\newcommand{\beann}{\begin{eqnarray*}}
\newcommand{\eeann}{\end{eqnarray*}}
\newcommand{\bes} {\begin{subequations}}
\newcommand{\ees} {\end{subequations}}

\newcommand{\ignore}[1]{}

\newcommand{\rmd}{{\text d}}

\begin{document}

\title{Discriminating Non-Isomorphic Graphs with an Experimental Quantum Annealer}

\author{Zoe Gonzalez Izquierdo}
\email{zgonzale@usc.edu}
\affiliation{Department of Physics and Astronomy, and Center for Quantum Information Science \& Technology, University of Southern California, Los Angeles, California 90089, USA}
\affiliation{Information Sciences Institute, University of Southern California, Marina del Rey, California 90292, USA}
\author{Ruilin Zhou}
\affiliation{Department of Electrical and Computer Engineering, Northwestern University, Evanston, Illinois 60208, USA}

\author{Klas Markstr\"{o}m}
\affiliation{Department of Mathematics and Mathematical Statistics, Ume\r{a} University, Ume\r{a}, Sweden}

\author{Itay Hen}
\affiliation{Department of Physics and Astronomy, and Center for Quantum Information Science \& Technology, University of Southern California, Los Angeles, California 90089, USA}
\affiliation{Information Sciences Institute, University of Southern California, Marina del Rey, California 90292, USA}

\begin{abstract}
\noindent We demonstrate experimentally the ability of a quantum annealer to distinguish between sets of non-isomorphic graphs that share the same classical Ising spectrum.  
Utilizing the pause-and-quench features recently introduced into D-Wave quantum annealing processors, which allow the user to 
probe the quantum Hamiltonian realized in the middle of an anneal, we show that obtaining thermal averages of diagonal observables of `classically indistinguishable' non-isomorphic graphs encoded into transverse-field Ising Hamiltonians enable their discrimination.
We discuss the significance of our results in the context of the graph isomorphism problem.
\end{abstract}

\maketitle

\section{Introduction}
\label{sec:intro}

Commercial quantum annealers~\cite{harris2010, harris2010_2, johnson2011} irrupted into the quantum computing scene about a decade ago, bringing along great promise---and just as much skepticism~\cite{vandam2002, znidaric2006}. Whether these machines do or could in principle provide any advantages over their classical analogues---or whether they could even be considered genuine quantum processors~\cite{boixo2013, boixo2013_2, smolin2014, shin2014, Albash2015}---is still a topic of much debate~\cite{ronnow2014, MartinMayor2015, Mandra2017, Mandra2017_2}.

While concrete evidence illustrating their usefulness as practical solvers of optimization problems is still scarce~\cite{Hen2015, katzgraber2015} the possibility that they can function as efficient Boltzmann samplers~\cite{Amin2018, 2018Marshall,Li2019} or as effective quantum simulators~\cite{2018King, 2018Harris, King2019, Bando2020} has been gaining traction.

Alongside efforts to extend the range of potential applications for quantum annealers, improvements to these devices are continuously being made. Notably, currently available devices allow the user to exercise additional control over various annealing parameters~\cite{dwave_tech} and determine, to a certain extent, the schedule of the anneal, which until recently could only be set to interpolate directly between a transverse-field at the beginning of the anneal and a classical Ising spin glass at the end of it. In particular, novel pause-and-quench capabilities enable some flexibility in determining the rates at which the classical Ising and transverse field Hamiltonians shift their respective strengths. 
This capability of pausing---the temporary halting of the anneal midway---deployed shortly after the minimum gap, has already been demonstrated to increase the success rates of the quantum annealer when used as an optimizer~\cite{2018Marshall, Passarelli2019}, by giving the system time to thermalize after the minimum gap.

Combining said pause with a sufficiently fast quench to the end of the anneal, such that the evolution during the quench is highly non-adiabatic, allows the user to effectively measure the system at the time of the pause, rendering the annealer a quantum simulator, which provides access to thermal properties of quantum, rather than classical, Hamiltonians. This new ability has already been tested with varying degrees of success~\cite{2018King, 2018Harris, King2019, gonzalez20}.

Another novel feature that has recently been introduced into experimental quantum annealers is that of reverse annealing. Here, the initial Hamiltonian is set to be the Ising spin glass and the transverse field is ramped up from zero to some intermediate value, and then brought back to zero again. 
This capability has been shown to help with the exploitation of regions of interest in the solution space of various optimization problems~\cite{Venturelli2018, Ohkuwa2018}.



In this work, we examine the possibility that mid-anneal probing through pause-and-quench may be utilized toward accomplishing an altogether different enterprise, namely, to solve instances of the Graph Isomorphism problem---the task of deciding whether two graphs are isomorphic (i.e., identical up to vertex relabeling). The possibility of using quantum annealers in this manner was first proposed almost a decade ago~\cite{2012Hen,2014Gaitan} and later experimentally attempted with somewhat limited success on an older-generation experimental quantum device that lacked the pause-and-quench features~\cite{2014Vinci}. 

Here, we study the power of quantum annealers equipped with pause-and-quench capabilities to distinguish between non-isomorphic graphs mapped onto transverse-field Ising Hamiltonians. 
We find that while measurements at the end of the anneal have only limited discrimination capabilities, mid-anneal measurements allow the extraction of sufficient information from the spectra of strictly quantum Hamiltonians and are in turn powerful enough to distinguish all tested sets of non-isomorphic graphs.

The paper is organized as follows. In Sec.~\ref{sec:theory}, we provide some background on the Graph Isomorphism problem, and the means for solving instances of the problem using quantum annealers. Sec.~\ref{sec:new_methods} relays the technical details of our experiments, including information about the specific parameters used. Results are presented and discussed in Sec.~\ref{sec:results}, and conclusions and final remarks laid out in Sec.~\ref{sec:conclusions}.

\section{Discriminating between graphs using a quantum annealer} 
\label{sec:theory}

A graph $G = (V, E)$, is a pair of sets, with $V$ being the set of vertices and $E$ the set of edges (unordered pairs of vertices), such that $E \subseteq [V]^2$~\cite{Diestel}. We denote the number of vertices $|V|$ by $n$. Two graphs $G$ and $G'$ are said to be isomorphic, $G \cong G'$, when there exists a bijection between them that preserves vertex adjacency, that is, one can be converted to the other by a simple relabeling of its vertices. 

Determining whether two graphs are isomorphic is known as the Graph Isomorphism (GI) problem~\cite{gip_book}. The best general case algorithm for solving it is quasi-polynomial in graph size~\cite{2015Babai}. 
One way of solving the GI problem is by comparing graph invariants: maps that take a graph as their argument and assign equal values to isomorphic graphs. A specific method to construct graph invariants results from encoding the structure of the graph on an Ising Hamiltonian~\cite{2012Hen}. To do this, to every edge $(i, j)$ in the graph we assign a term $\sigma_{i}^{z} \sigma_{j}^{z}$ in the Hamiltonian, where $\sigma_i^z$ represents an individual Pauli-$z$ matrix on spin $i$. Optionally, we can add an external, homogeneous longitudinal field to the spins. We call this the `problem Hamiltonian' of the graph:
\begin{equation}
H_p(G) = \sum\limits_{(i,j) \in E(G)} \sigma_{i}^{z} \sigma_{j}^{z} + \sum\limits_{i=1}^{n} \sigma_{i}^{z}.
\label{eq:problem_h}
\end{equation}
The invariant of interest is the spectrum of $H_p$, and two graphs $G$ and $G'$ are said to be co-Ising if $H_p(G)$ and $H_p(G')$ have the same spectra. The spectrum of $H_p$ is however not a complete invariant, meaning that non-isomorphic, co-Ising graphs exist \cite{AM09}, and trying to use $H_p$ as a graph discriminator to solve the GI problem would therefore fail in the general case.

Quantum annealers offer the exciting possibility of extending the set of Hamiltonian graph invariants into the quantum regime. This can most simply be done by adding a quantum term, such as a homogeneous transverse field, to the classical Ising model onto which the graph is mapped. In this case, the Hamiltonian can be written as
\begin{equation}\label{Hamiltonian}
H(G, s) = A(s)H_d + B(s)H_p(G),
\end{equation}
where $s$ is a rescaled, dimensionless time parameter, $s \in [0, 1]$. $A(s)$ and $B(s)$ are functions of $s$ such that $A(0) \gg B(0) \simeq 0$ and $B(1) \gg A(1) \simeq 0$,
and $H_d$ is the standard transverse field driver Hamiltonian in quantum annealing:
\begin{equation}
H_d =  \sum\limits_{i=1}^{n} \sigma_{i}^{x}.
\end{equation}
While it is known that there are non-isomorphic graphs for which the spectra of their respective $H_p$ are identical, so far graphs with the same spectra for the full quantum Hamiltonian $H(G, s)$--- quantum co-Ising ---which are non-isomorphic are not known to exist. 

Calculating the full quantum Ising spectrum of a given graph requires in general the diagonalization of a $2^n \times 2^n$ matrix, which quickly becomes intractable as $n$ increases. Quantum annealers, which implement the above quantum Ising Hamiltonian $H(G, s)$, could alternatively be used to extract valuable information without resorting to calculating its entire spectrum~\cite{2012Hen}. This can be done if the annealing process is stopped midway at some intermediate $s$ value, $0 < s <1$, and measurements are then taken in order to obtain an estimation of thermal averages of physical observables. 

Experimentally testing the conjecture that physical quantum annealers can distinguish non-isomorphic graphs poses however two main challenges. The first is that currently available processors only allow diagonal measurements to be performed, restricting to some extent the amount of information accessible to the experimenter. 
Moreover, to ensure that the quantum annealer identifies isomorphic graphs as such, one must also choose to measure quantities that remain unchanged under vertex relabeling. 

In this study, we measure the classical energy $E \equiv \langle H_p \rangle$, the $z$-magnetization $\langle M_z \rangle$, the spin-glass order parameter~\cite{2012Hen},
\begin{equation}
    Q_2 = \sqrt{\frac{1}{N(N-1)}\sum_{i \neq j} \langle \sigma_i^z \sigma_j^z \rangle^2}\,,
\end{equation}
and the next-nearest neighbor interaction energy~\cite{2014Vinci}
\begin{equation}
    \Omega^2 = \sum_{i,j} [A^2(G)]_{i,j} \sigma_i^z \sigma_j^z.
\end{equation}

\begin{figure}[h]
    \centering
    \includegraphics[width=\columnwidth]{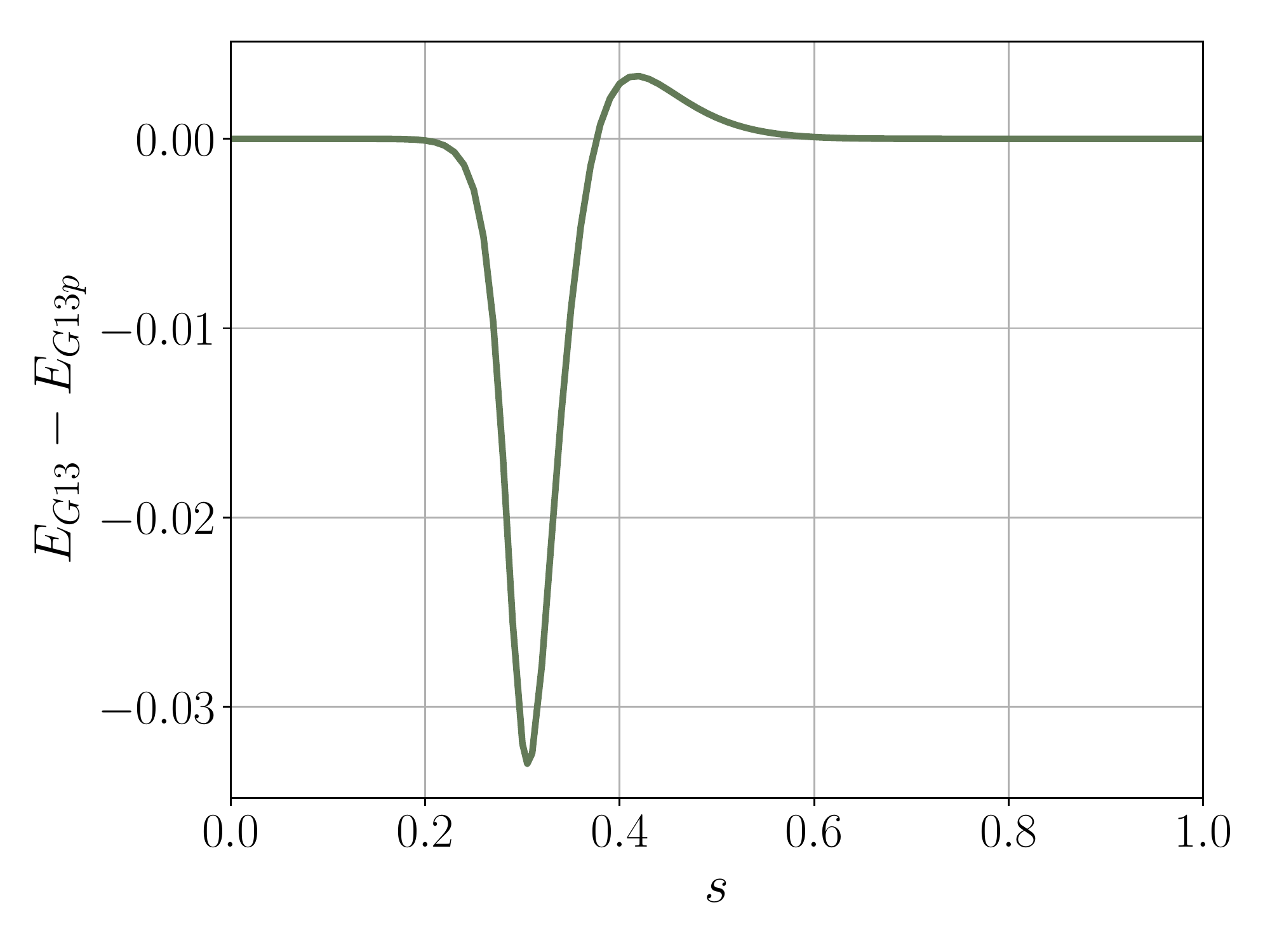}
    \caption{Difference in classical energy $E = \langle H_p \rangle$ throughout the anneal at $T=12$mK (device operating temperature) for the smallest pair of graphs we test, $G13$ and $G13p$. 
    }
    \label{fig:g13_exact_energy}
\end{figure}

The second challenge that current quantum annealing technology faces is that it only allows measurements to be taken at the end of the anneal, where only the classical $H_p$ component has non-zero strength. This setting is appropriate for the standard usage of a quantum annealer as an optimizer, when the objective is finding the ground state of $H_p$, and the transverse field is just a means to arriving at it. However, as was discussed above, for distinguishing between non-isomorphic graphs, sampling from a thermal state of a classical Hamiltonian is insufficient in general. To illustrate why this is so, consider Fig.~\ref{fig:g13_exact_energy}, which shows 
 \hbox{$\Delta E = E_{G13} - E_{G13p}$} for one of the pairs of graphs we try to distinguish, denoted $G13$ and $G13p$, with $n=13$ (small enough to be directly diagonalized). These graphs are indistinguishable at the end of the anneal as evidenced by the $\Delta E = 0$ at $s=1$ (as well as in the $s=0$ limit when their Hamiltonians are both only the transverse field), and the difference between them is pronounced enough only well within the quantum regime. A measurement near $s=0.3$, where $\Delta E$ peaks, would thus be ideal for distinguishing them. 
The inability of sampling from a classical Hamiltonian to discriminate non-isomorphic graphs was also demonstrated in Ref.~\cite{2014Vinci}. There, the authors tested whether an experimental quantum annealer could lead to distinguishable statistics when measuring only at the end of the anneal, concluding this cannot be done in general.


While currently available hardware still only allows $s=1$ measurements, mid-anneal differences may be captured thanks to the newly added capability of employing fast quenches~\cite{2018Marshall, 2018Ohkuwa, 2018King, 2018Harris}.
These in turn induce highly non-adiabatic transitions. While still far from being the ideal instantaneous quench (which is equivalent to a mid-anneal measurement), fast quenches allow, to some extent, access to information about the spectra of the quantum graph Hamiltonians. 
As we demonstrate below, combining these fast quenches with reverse annealing and pausing can be used to substantially increase the distinguishability power of quantum annealers.

\section{Methods}
\label{sec:new_methods}
The quantum annealing processor we use in our experiment is the D-Wave 2000Q quantum annealer located at NASA Ames Research Center. Released in 2016, its 2048 qubit chip has a $16 \times 16$-cell Chimera connectivity. Chimera graphs of size $s$ ($C_s$), consist of $s^2$ Chimera cells arranged in a square pattern. Each Chimera cell is a complete bipartite graph with 8 vertices, or $K_{4,4}$. Each vertex is connected to 4 others within the cell, and also to 2 more outside the cell (except for cells on the boundaries). Each edge can be assigned a coupling strength $J_{ij}$ and each qubit a longitudinal field $h_i$.

Every graph $G$ we test is mapped to a problem Hamiltonian $H_p$ as given in Eq.~(\ref{eq:problem_h}).
Each graph vertex is represented by a qubit on the chip, and each edge of the graph is assigned the appropriate coupler with strength $J_{ij}=1$ (non-edge couplers are set to 0). The particular choice of physical qubits and couplers representing the logical graph is what we refer to as embedding. In general, there will be numerous possible embedding choices for each graph. In addition, we apply a homogeneous external longitudinal field on all qubits, with \hbox{$h_i = 1 \; \forall i$}. We also perform gauge averaging to mitigate potential intrinsic biases in the physical components. This is done by repeating each run several times, each applying a different random transformation to the couplers and longitudinal fields of the form $J_{ij} \rightarrow a_i a_j J_{ij}$, $h_i \rightarrow a_i h_i$, with $a_i \in \{ -1, +1 \}$. These are unitary transformations that leave the energy spectrum unchanged, and the solution to the original problem can be easily recovered from that of the gauge-transformed problem by reversing the transformation.

\begin{figure}[htp]
    \centering
    \includegraphics[width=\columnwidth]{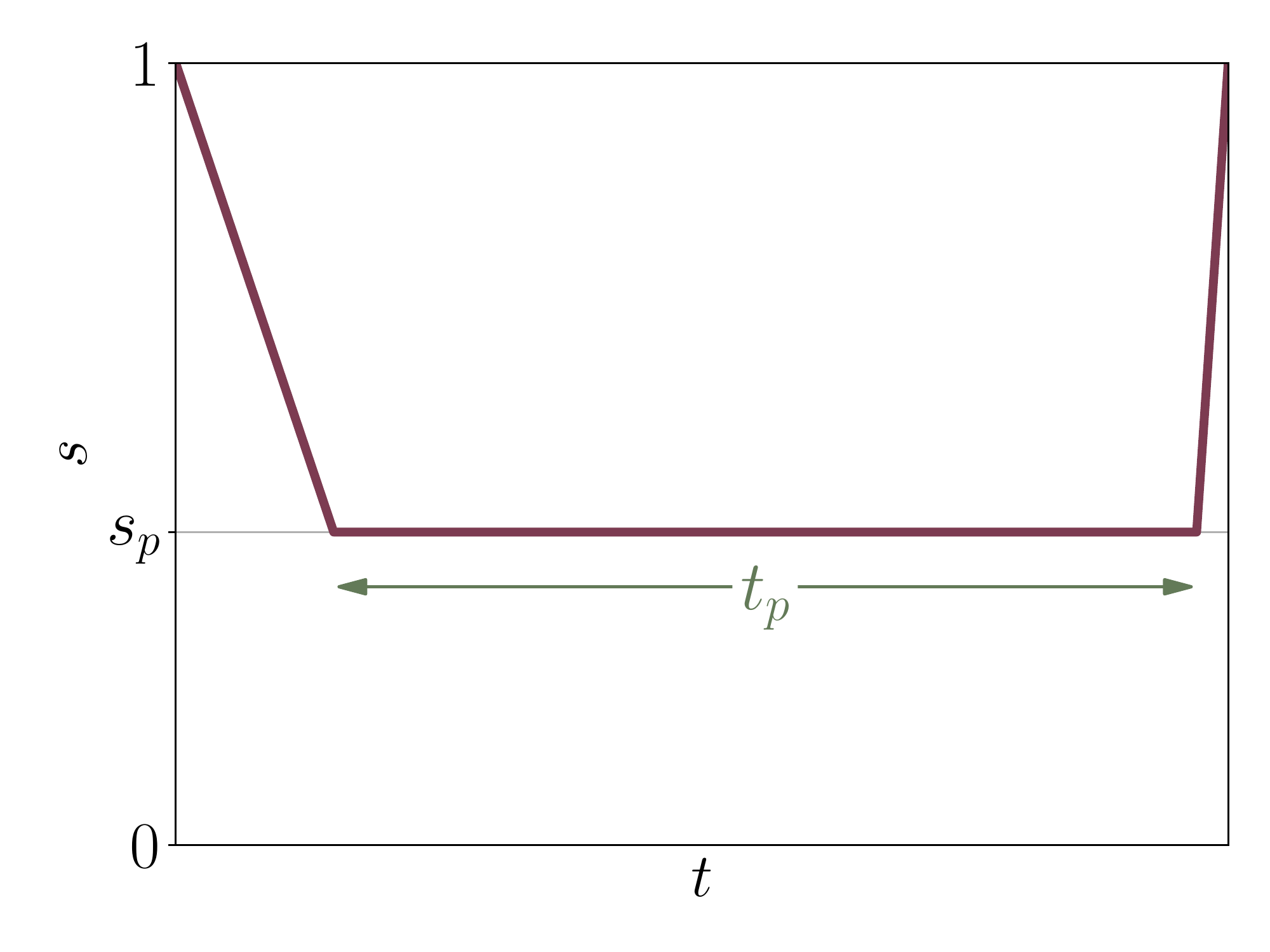}
    \caption{Diagram of the annealing schedule $s(t)$ for reverse annealing with a pause and a final quench (not to scale). The system is initialized at $s=1$, annealed back to an intermediate $s_p$, paused for a time $t_p$ and finally annealed to $s=1$ again, this time at the maximum rate.}
    \label{fig:s_diagram}
\end{figure}

In our annealing procedure, we employ a reverse annealing protocol with a pause, followed by a quench. With reverse annealing, $s(t=0)=1$ and the initial Hamiltonian is $H(s=1) \approx H_p$. The system is prepared in a classical state set by the user (in our case, all spins up), and the anneal starts `in reverse' (i.e. decreasing $s$) to some intermediate value $s=s_p$. The same path must then be traced back, as the process is required to end at $s=1$. In our case we also choose to introduce a pause of length $t_p$ at $s_p$. A qualitative depiction of our schedule is shown in Fig.~\ref{fig:s_diagram}. After the first anneal is completed, one can choose to either reinitialize to the chosen classical state, or take the final state for each anneal as the initial state for the next. We choose the latter option.

In our runs, the duration of the reverse part of the anneal as well as the pause are set constant (at $1\mu$s and $100\mu$s,  respectively), as is the rate of the quench (the fastest possible, $1\mu$s$^{-1}$). The location of the pause $s_p$ is varied in $[0.1, 0.2, \dots , 0.9]$ to obtain statistics at different points in the anneal.
This schedule is designed to allow us to tap into the quantum spectrum of the Hamiltonian at the time of the pause $s_p$, by letting the system relax to the thermal state of $H(s_p)$ while also trying to prevent further thermalization during the quench. Rather than obtaining the ground state (or thermal distribution) of $H_p$ as we would for optimization purposes, here we are interested in obtaining configurations whose statistics retain information about the state of the system at $s_p$ where the Hamiltonian is non-classical. 

Recent results~\cite{gonzalez20} suggest that the current fastest rate for the quench is, in general, not sufficiently fast to fully prevent the state from changing during the quest, and thus measurements might not faithfully represent the thermal state at $H(s_p)$. For our purpose, the requirement that the state remains unchanged during the quench is however not strictly necessary. We only require that enough information is preserved during the quench so that the differences between quantum spectra are not completely washed out by the time a measurement is taken at the end of the anneal.


For each graph and pause location, we choose 200 random gauges and perform 1000 anneals per gauge, obtaining 200,000 final configurations per set of parameters. Diagonal observables---classical energy $\langle H_p \rangle$, magnetization $\langle M_z \rangle$, spin-glass order parameter $Q_2$ and next-nearest neighbor interaction $\Omega^2$ (defined in Sec.~\ref{sec:theory})---are calculated for each gauge, and then a bootstrap over the gauges is performed and the mean and 95\% confidence interval values reported.

To ensure that the annealer does not falsely distinguish isomorphic graphs, and that differences in results stem in fact from non-isomorphicity, rather than biases or noise, we test isomorphic graphs as well, as a baseline. 
This is done in one of two ways: in the case of pairs of graphs that can be embedded using the same qubits, a third graph, also defined on the same set of qubits and isomorphic to the first, is tested alongside the pair. Isomorphic variants are easily generated by random relabelings of graph vertices.
We note that since the choice of embedding is limited by the rigid, sparse connectivity of the Chimera layout, embedding on the same qubits is not always possible. When that is the case, we do not compare a single embedding choice for each graph, due to the fact that different physical qubits and couplers might suffer from intrinsic biases. Instead, each graph is assigned several different embeddings, and results are averaged over those.
We find that the differences between non-isomorphic graphs remain statistically significant even averaging over several different embeddings. 

\section{Results}
\label{sec:results}

To test the extent to which quantum annealing processors can discriminate `classically indistinguishable' non-isomorphic graphs, we consider several sets of graphs of varying sizes (ranging between $n=13 $ and $n=33$ vertices). These graphs are specially constructed to be classically co-Ising and native to the Chimera architecture  (the reader is referred to the Appendix for details on the construction process and the particulars of the graphs tested).
\begin{figure}[htp]
    \centering
    \includegraphics[width=0.93\columnwidth]{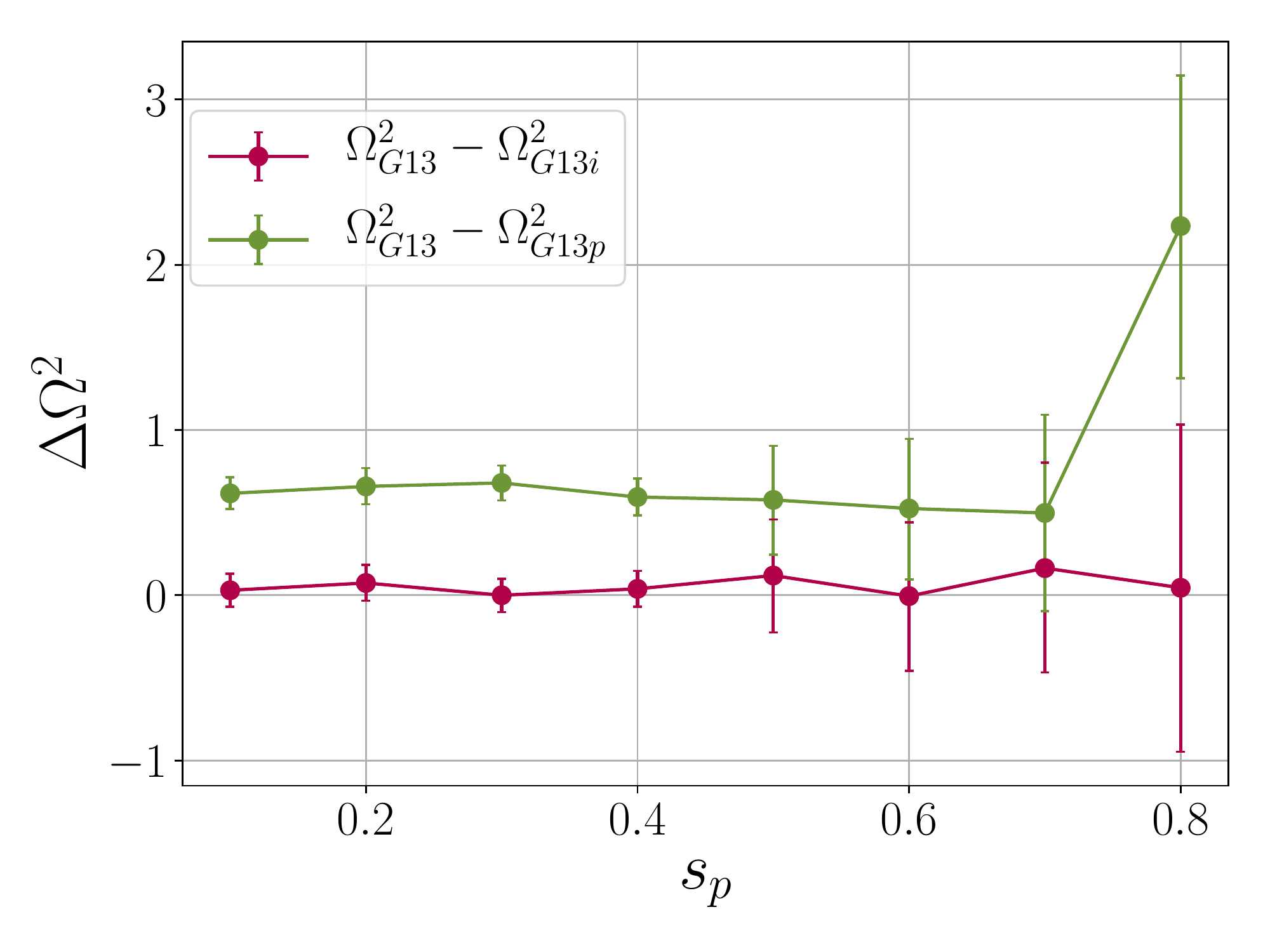}
    \includegraphics[width=0.95\columnwidth]{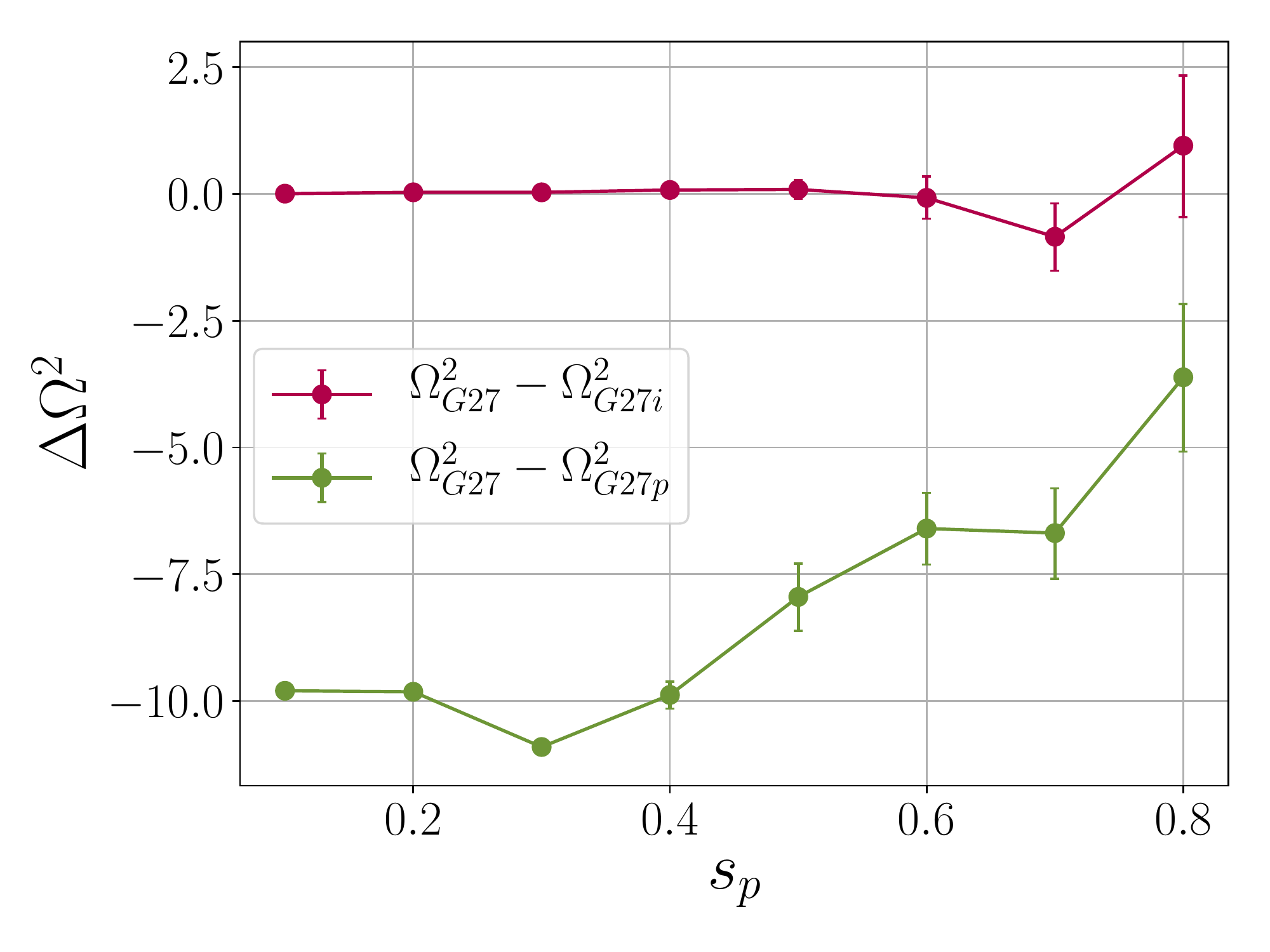}
    \caption{Differences in next-nearest neighbor interaction energy as a function of the pause point $s_p$. $\langle \Omega^2 \rangle$ suffices to discriminate between $G13$ and $G13p$ (top) and between $G27$ and $G27p$ (bottom). The red line shows the difference between the two isomorphic graphs $G13$ and $G13i$ (top) and $G27$ and $G27i$ (bottom), staying relatively constant around 0, while the green line represents the difference between the non-isomorphic graphs $G13$ and $G13p$ (top) and $G27$ and $G27p$ (bottom), which remains non-zero throughout. Error bars show the 95\% confidence interval after performing a bootstrap over the 200 gauges.}
    \label{fig:g13_27}
\end{figure}
Importantly, to verify that the experimental annealer does not falsely distinguish isomorphic graphs as well, we add to each set of non-isomorphic graphs we wish to discriminate isomorphic variants of some of the graphs. When the set is a pair that can be embedded using the same set of qubits, two sets of data are presented for each observable: the difference in the observable between the two isomorphic graphs---which should be zero (up to statistical errors) if the annealer sees them as such---and the difference in the observable between the non-isomorphic graphs---which indicates that the annealer tells them apart if the value is non-zero. When the graphs within a set cannot be embedded using the same set of qubits, we average over equivalent embeddings. To successfully distinguish non-isomorphic graphs, differences in the observables must be larger than the variation across embeddings. We find that in the regions where differences are most pronounced, the difference in observable averages is typically an order of magnitude greater than the error bars which represent variations across embeddings. We next report our main findings. 

We first discuss the two pairs of graphs previously tested in Ref.~\cite{2014Vinci}. There, an $n=13$-vertex pair of graphs was distinguished by examining sorted individual values of a triplet of observables, while for the larger pair (with $n=27$) the two graphs could not be told apart. Using a mid-anneal pause, which allows us to probe quantum Hamiltonians in the middle of the anneal, we are able to measure the differences in the quantum spectra for both pairs of graphs, and to do so simply from the outcome of a single diagonal observable, such as $\langle H_p \rangle$, $Q_2$, $\langle M_z \rangle$ or $\langle \Omega^2 \rangle$. 

The above pairs of graphs have the advantage that they could be embedded using the same set of qubits on the annealer hardware. To ensure that we are in fact measuring differences in quantum spectra, rather than differences caused by noise or other factors, we also test a third graph alongside each pair, that is isomorphic to one of the two. We find the quantum annealer is unable to `distinguish' between the isomorphic graphs, as should be the case. Plots depicting the measured differences for the $n=13$ pair and for the previously indistinguishable $n=27$ pair are shown in Fig.~\ref{fig:g13_27}. 

We next examine new tuples of graphs designed to be classically co-Ising, native to the Chimera architecture, as detailed in the Appendix, and have the added difficulty that they cannot in general be embedded using the same sets of qubits.  We examine in particular a pair with $n=17$ vertices and two sets of four graphs of sizes $n=25$ and $n=33$. 

\begin{figure*}[ht]
    \centering
    \includegraphics[width=0.65\columnwidth]{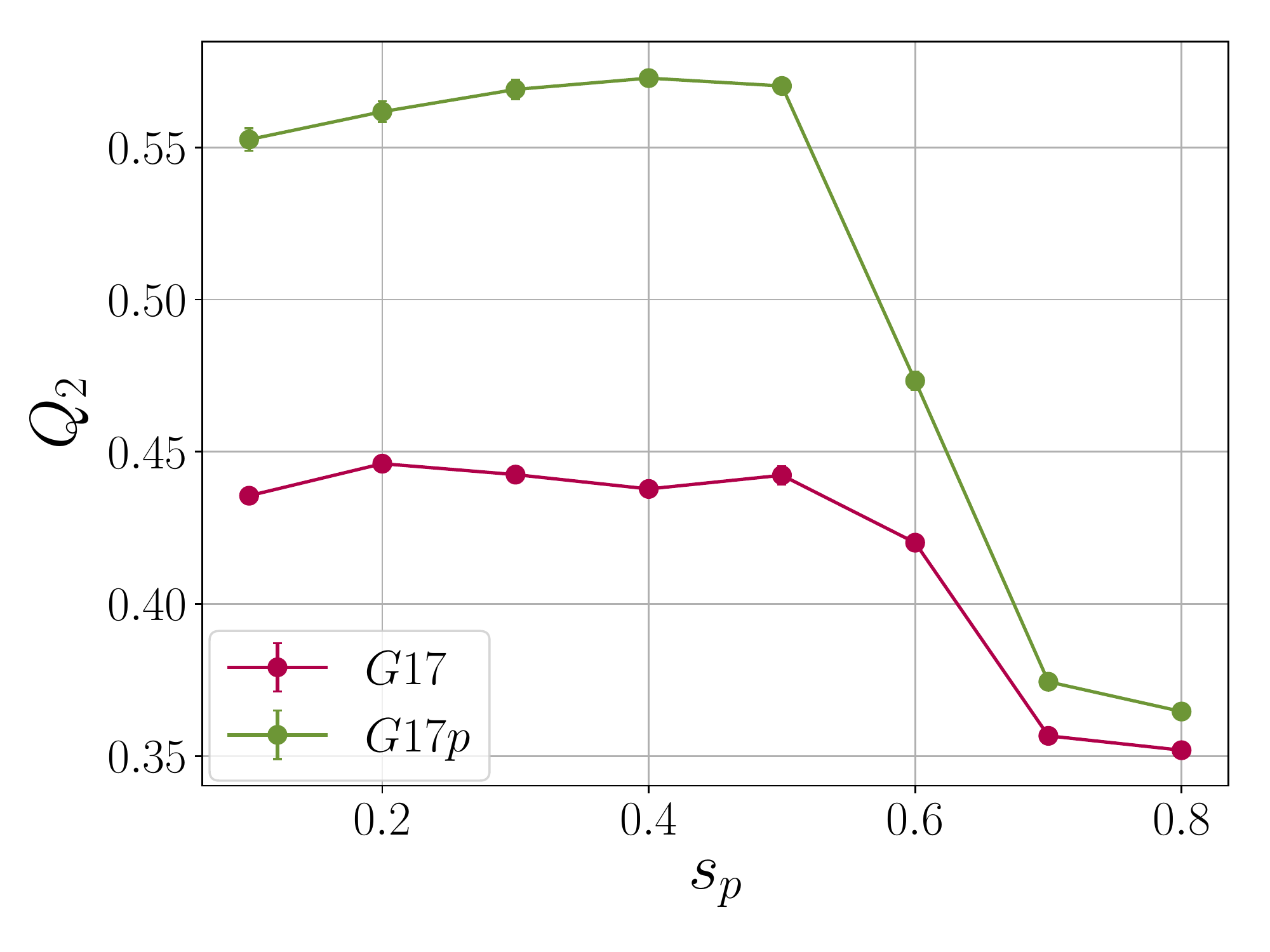}
    \includegraphics[width=0.65\columnwidth]{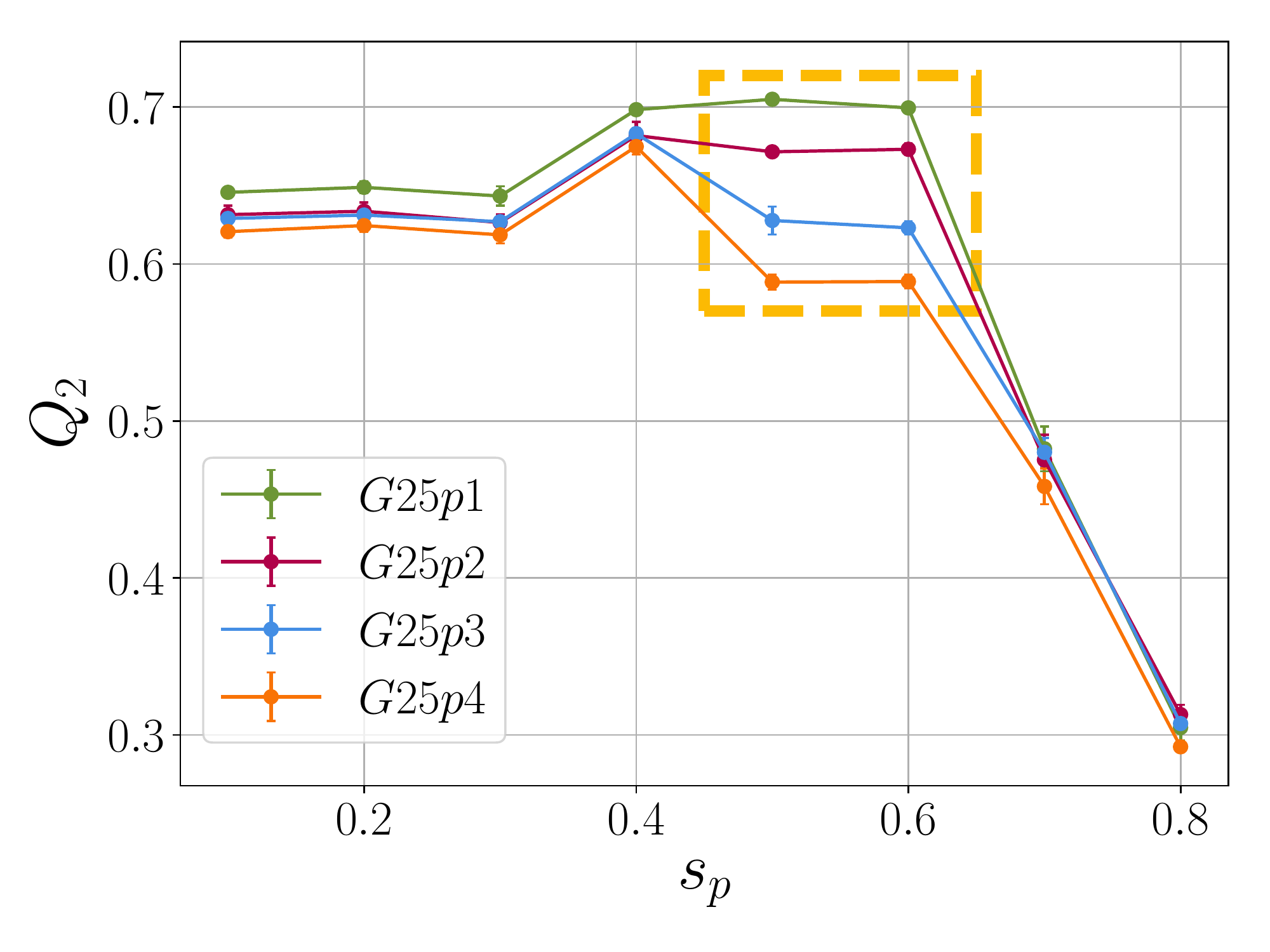}
     \includegraphics[width=0.64\columnwidth]{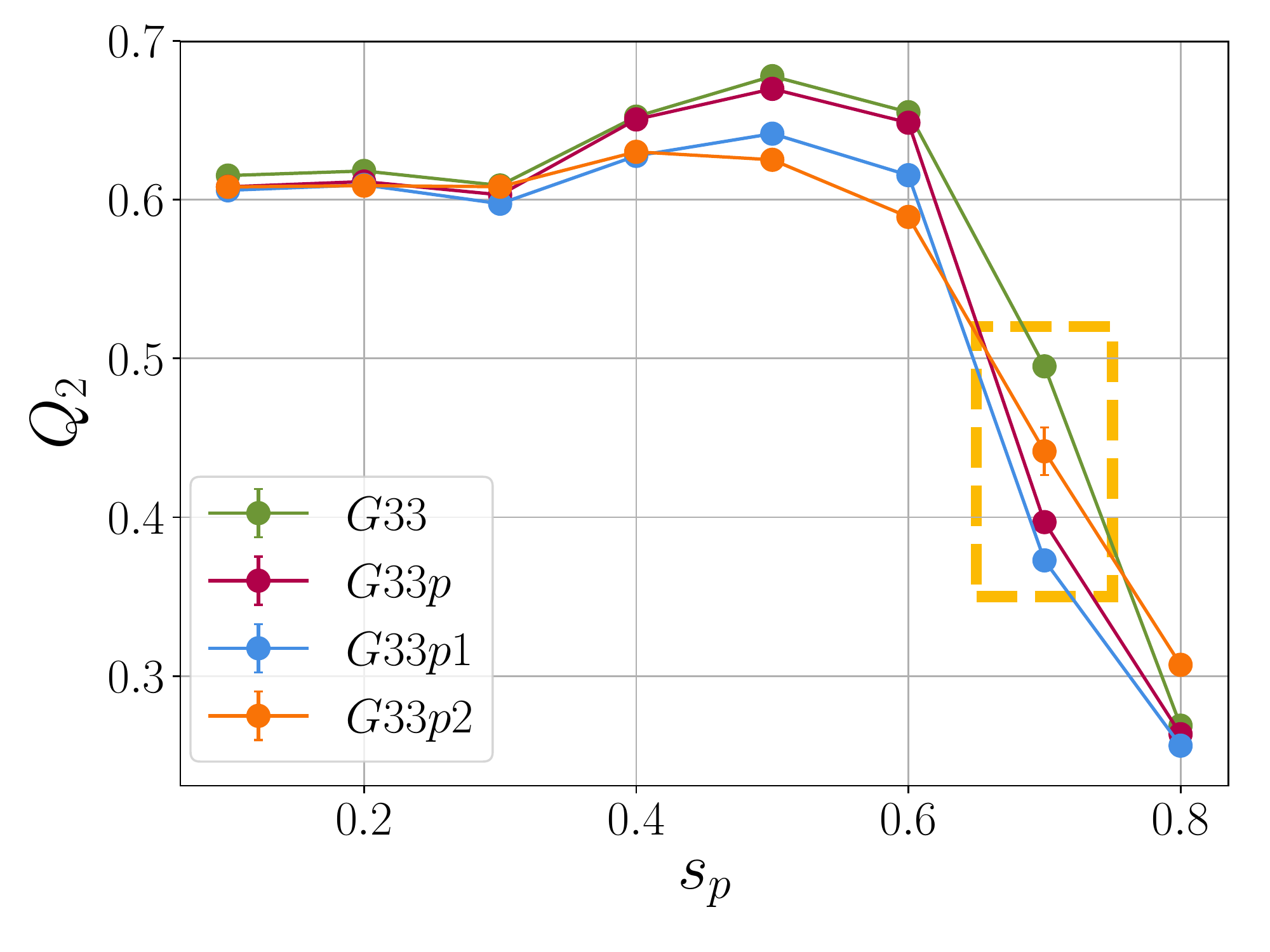}
    \caption{\textbf{Left:} The spin-glass order parameter $Q_2$ as a function of pause point $s_p$. Measurements of $Q_2$ can distinguish between $G17$ and $G17p$. Data points are averaged over 5 different embeddings, with error bars showing the standard error.
\textbf{Middle:} A group of four classically equivalent graphs with $n=25$ becomes distinguishable when probing the quantum regime mid-anneal. If the pause is too early or too late, we are not able to measure the differences across these four graphs. Each data point is averaged over 5 different embeddings, with error bars showing the standard error of the mean. Individual embedding results were first calculated using a bootstrap over the 200 gauges. The differences between the four graphs are most pronounced in the range $s \in [0.5,0.6]$ (marked by the dashed contour).
   \textbf{Right:} Similar to the $n=25$ tuple, $Q_2$ can distinguish all four $G33$ graphs with a pause around $s_p=0.7$.}
    \label{fig:g17_25_33}
\end{figure*}

To confirm that any measured differences are in fact due to non-isomorphicity, we run several different embeddings for each of these graphs, and average the measurement outcomes over the embeddings of each graph. Ideally, we expect to obtain similar outcomes for different embeddings of same graphs. Indeed, this is what we find. 
Figure~\ref{fig:g17_25_33} (left) depicts the distinguishability power of the experimental quantum annealer for the $n=17$ pair of graphs. Here, we pick the spin-glass order parameter $Q_2$ as the discriminator of choice.   
In Fig.~\ref{fig:g17_25_33} (middle) we show how, for a $4$-tuple of graphs of size $n=25$, pausing in a region in the middle of the anneal brings out the differences in the quantum spectra between the four graphs, which are much larger than the small deviations due to the different embeddings.
We similarly test another 4-tuple with $n=33$ (Fig.~\ref{fig:g17_25_33} right).

As is evident from these figures, differences between the tested sets of non-isomorphic graphs are most pronounced in the region where $s$ is far from the ends of the anneal, well into the `quantum regime'. In the limit of $s \approx1$, no differences are present as our graphs are specially constructed to be classically co-Ising. In the other limit $s \approx 0$, no differences should be detected because only the $H_d$ component, which is identical across all graphs, has non-zero strength. Interestingly, we find that in practice, as is evident from the bottom panel of Fig.~\ref{fig:g13_27}, outcomes from the non-isomorphic graphs can be quite different even when the pause is rather close to $s=0$.
One plausible explanation for the above observation is that for pauses that take place at small $s$ values, the quench to $s=1$ cannot be performed fast enough to preserve the state of the system at the pause point. This is expected since for small $s$ values the quench has to pass through the minimum gap, in the vicinity of which thermalization processes take place at very rapid rates~\cite{2018Marshall}. 

\section{Conclusions and outlook}
\label{sec:conclusions}

We presented results demonstrating the ability of an experimental quantum annealer to solve instances of the graph isomorphism problem. We illustrated that a physical quantum annealer's ability to distinguish between `classically indistinguishable' graphs---graphs whose classical Ising spectra are identical---may be accomplished by utilizing pause-and-quench features, which allow the user to probe the thermal properties of strictly quantum Hamiltonians; capabilities that have so far been out of reach for standard annealing processors.
We also showed that while measurements at the end of the evolution following a standard annealing protocol are not able to distinguish between non-isomorphic graphs, introducing a mid-anneal pause followed by a fast quench 
allowed us to access properties of strictly quantum Hamiltonians, providing the annealer the necessary distinguishability power that standard annealing protocols lack. 

We confirmed the robustness of our technique across different graph choices and native embeddings, as a crucial first step in the quest to expand the scope of this method. With the advent of new architectures with increased connectivity that will soon succeed the Chimera architecture of current D-Wave devices, the pool of native graphs is expected to expand significantly. For larger, more connected graphs, being able to rely on heuristic methods for embedding-finding will likely be necessary, and the success of this method relies on the consistency of results across embeddings.

We found that despite pause-and-quench being far from a perfect mechanism to substitute a measurement in the middle (as quench times cannot be made arbitrarily short and pause times cannot be made arbitrarily long)~\cite{gonzalez20}, the method we employed performed as intended insofar as extracting sufficient information about the differences in the quantum spectra of non-isomorphic graphs at the time of the pause. We expect that further improvements in technology, e.g., a faster quench or, better yet, novel mechanisms for taking measurements in the region $0<s<1$, would lead to results that capture the quantum spectrum more faithfully, enhancing the annealer's discriminating capabilities. Nonetheless, the current setup proved sufficient to distinguish all tested sets of graphs by estimating thermal averages of one or several physical (diagonal) observables, where at least one of them---and usually more---show statistically significant differences between the non-isomorphic graphs. This brings to fruition the idea, proposed almost a decade ago~\cite{2012Hen}, that classically co-spectral graphs could be distinguished by a quantum annealer via measurements of their quantum properties.

Importantly, in this work we have not discussed the time resources needed for quantum-assisted graph discrimination, nor did we discuss the potential of observing quantum speedups, which remains an interesting consideration to be addressed in future work. While the present study focuses on feasibility, a rigorous investigation of performance and its scaling with the size of the graphs will be necessary to test this method against classical algorithms. Differences between reverse and forward annealing with a pause should be considered, including the duration of the pause, as well as other annealing parameters, adjusted to minimize time-to-solution.

\begin{acknowledgements}
This research is based upon work (partially) supported by the Office of
the Director of National Intelligence (ODNI), Intelligence Advanced
Research Projects Activity (IARPA), via the U.S. Army Research Office
contract W911NF-17-C-0050. 
Klas Markstr\"{o}m was supported by The Swedish Research Council grant 2014-4897.
The views and conclusions contained herein are
those of the authors and should not be interpreted as necessarily
representing the official policies or endorsements, either expressed or
implied, of the ODNI, IARPA, or the U.S. Government. The U.S. Government
is authorized to reproduce and distribute reprints for Governmental
purposes notwithstanding any copyright annotation thereon.
\end{acknowledgements}

\bibliography{graph_iso_ref.bib}

\appendix

\section*{Appendix: Choice of Graphs}

We discuss the construction of pairs (or larger tuples) of non-isomorphic graphs that are co-Ising, i.e., have identical classical Ising spectra. 

We first define the Ising polynomial, or partition function for the Ising model~\cite{AM09}:
\begin{equation}
    Z(G, x, y) = \sum_{\vec{s} \in \Omega} x^{E(\vec{s})} y^{M_z(\vec{s})},
\end{equation}
where $\vec{s}$ is a spin configuration, $\Omega$ is the space of all possible configurations, and $E(\vec{s})$ and $M_z(\vec{s})$ are the energy and magnetization of the configuration $\vec{s}$. Co-Ising graphs have the same Ising polynomial.

For small $n$ ($\leq 10$), a full classification of equivalence classes of co-Ising graphs is known. The first such examples were given in Ref.~\cite{AM09}, where they were found via a complete search of all small graphs, and  the search was extended in Ref.~\cite{M17}, where non-isomorphic graphs which have the same partition function for all two-state spin models were found.  But due to the rapidly growing number of non-isomorphic graphs, even when restricted to e.g. regular graphs, and the rarity of co-Ising pairs, any complete search will be limited to quite small sizes.

In Ref.~\cite{AM09} a technique was introduced for constructing large families of non-isomorphic graphs with the same Ising-polynomial.
This method was extended in Ref.~\cite{MW10} to the random-cluster model.  Using data from Refs.~\cite{AM09,M17} we use a simple version of this method to construct larger graphs to use as benchmarks.

This method makes use of the \emph{rooted Ising-polynomial} $Z(G,v,x,y)$ of a graph $G$. Given a vertex $v$  in the graph $G$ this polynomial is defined as the usual Ising-polynomial except that we only sum over those states on $G$ in which the spin of $v$ is $+1$.  Note that the full Ising-polynomial is given by $Z(G,x,y)=Z(G,v,x,y)+Z(G,v,x,1/y)$.  
So,  if two graphs $G_1, G_2$ have the same rooted polynomials $Z(G_1,v_1,x,y)$ and $Z(G_2,v_2,x,y)$ for some vertex $v_1\in G_1$ and $v_2 \in G_2$ then they will have the same Ising-polynomial, but two graphs may have the same Ising-polynomial even if no such pair of vertices exist. 

The rooted polynomial has the useful property that if we start with any two graphs, and root vertices $v_1$ and $v_2$, and  build a new graph $G_3$ by identifying the vertices $v_1$ and $v_2$ into a new vertex $v_3$ then $$Z(G_3,v_3,x,y)=Z(G_1,v_1,x,y)Z(G_2,v_2,x,y)/y.$$
So, if we know the rooted polynomial for the smaller graphs $G_1$ and $G_2$ we can easily find the full Ising-polynomial of $G_3$, and if we start  with two different pairs of non-isomorphic graphs with the same rooted polynomials we can build two larger graphs which are Co-Ising, but which may not be isomorphic.  Similarly we can use more than two graphs in this product, as long as we divide by the right power of $y$. 

Using the examples from Ref.~\cite{M17} we search for pairs of graphs with the same rooted polynomials. Several such pairs exist, and we opt for a few pairs of tree-like starting graphs to keep them native to the Chimera architecture. While minor embedding~\cite{2008Choi} is a commonly used technique for dealing with non-native graphs, where several physical qubits linked by strong ferromagnetic couplings are used to represent a single node from the original graph, we avoid it for the present study.

All the tested graphs are listed below:

\begin{align*}
    G13 = [&(1, 8), (1, 10), (1, 11), (1, 13), (2, 9), (2, 11),\\
    & (2, 13), (3, 10), (3, 13), (4, 10), (5, 11), (6, 12), \\
    &(7, 12), (9, 12), (12, 13)] \\
    G13p = [&(1, 8), (1, 10), (1, 11), (1, 13), (2, 9), (2, 11), \\
    &(2, 13), (3, 10), (3, 11), (4, 10), (5, 12), (6, 12), \\
    &(7, 13), (8, 12), (12, 13)]
\end{align*}
\begin{align*}
    G17 = [&(1, 2), (1, 3), (1, 4), (1, 5), (4, 6), (4, 7), \\
    &(5, 8), (5, 9), (5, 10), (6, 11), (10, 12), (10, 13), \\
    &(10, 14), (11, 15), (11, 16), (12, 17)]\\
    G17p = [&(1, 2), (1, 3), (1, 4), (1, 5), (4, 6), (5, 7), \\
    &(5, 8), (5, 9), (6, 10), (6, 11), (9, 12), (9, 13), \\
    &(9, 14), (10, 15), (12, 16), (12, 17)]
\end{align*}
\begin{align*}
    G25p1 = [&(1, 6), (1, 7), (3, 7), (4, 8), (4, 9), (5, 8), \\
    &(5, 9), (6, 9), (10, 14), (10, 15), (11, 15), (12, 16), \\
    &(12, 17), (13, 16), (13, 17), (14, 17), (18, 22), \\
    &(18, 23), (19, 23), (20, 24), (20, 25), (21, 24), \\
    &(21, 25), (22, 25), (2, 6), (2, 14), (2, 22)] \\
    G25p2 = [&(1, 6), (1, 7), (3, 7), (4, 8), (4, 9), (5, 8), \\
    &(5, 9), (6, 9), (10, 14), (10, 15), (11, 15), (12, 16), \\
    &(12, 17), (13, 16), (13, 17), (14, 17), (18, 23), \\
    &(18, 25), (19, 22), (19, 24), (20, 23), (20, 25),\\
    &(21, 24), (21, 25), (2, 6), (2, 14), (2, 18)] \\
    G25p3 = [&(1, 6), (1, 7), (3, 7), (4, 8), (4, 9), (5, 8), \\
    &(5, 9), (6, 9), (10, 15), (10, 17), (11, 14), (11, 16), \\
    &(12, 15), (12, 17), (13, 16), (13, 17), (18, 23), \\
    &(18, 25), (19, 22), (19, 24), (20, 23), (20, 25), \\
    &(21, 24), (21, 25), (2, 6), (2, 10), (2, 18)] \\
    G25p4 = [&(1, 7), (1, 9), (2, 6), (2, 8), (3, 7), (3, 9), \\
    &(4, 8), (4, 9), (10, 15), (10, 17), (11, 14), (11, 16), \\
    &(12, 15), (12, 17), (13, 16), (13, 17), (18, 23), \\
    &(18, 25), (19, 22), (19, 24), (20, 23), (20, 25), \\
    &(21, 24), (21, 25), (5, 1), (5, 10), (5, 18)]
\end{align*}
\begin{align*}
    G27 = [&(1, 14), (1, 17), (2, 14), (2, 22), (3, 4), (3, 5), \\
    &(4, 10), (4, 12), (5, 11), (5, 13), (6, 7), (6, 8), \\
    &(6, 15), (7, 10), (7, 11), (8, 12), (8, 13), (9, 12), \\
    &(9, 13), (9, 14), (10, 15 ), (11, 15), (14, 15), (16, 17), \\
    &(16, 21), (17, 18), (18, 19), (19, 20), (20, 21), (22, 23), \\
    &(22, 27), (23, 24), (24, 25), (25, 26), (26, 27)] \\
    G27p = [&(1, 14), (1, 17), (2, 14), (2, 23), (3, 4), (3, 5), \\
    &(4, 10), (4, 11), (5, 12), (5, 13), (6, 7), (6, 8), \\
    &(6, 15), (7, 10), (7, 12), (8, 11), (8, 13), (9, 12), \\
    &(9, 13), (9, 14), (10, 15), (11, 15), (14, 15), (16, 17), \\
    &(16, 21), (17, 18), (18, 19), (19, 20), (20, 21), (22, 23), \\
    &(22, 27), (23, 24), (24, 25), (25, 26), (26, 27)]
\end{align*}
\begin{align*}
    G33 = [&(1, 6), (1, 7), (3, 7), (4, 8), (4, 9), (5, 8), \\
    &(5, 9), (6, 9), (10, 14), (10, 15), (11, 15), (12, 16), \\
    &(12, 17), (13, 16), (13, 17), (14, 17), (18, 22), (18, 23), \\
    &(19, 23), (20, 24), (20, 25), (21, 24), (21, 25), (22, 25), \\
    &(26, 30), (26, 31), (27, 31), (28, 32), (28, 33), (29, 32), \\
    &(29, 33), (30, 33), (2, 6), (2, 14), (2, 22), (2, 30)] \\
    G33p = [&(1, 6), (1, 7), (3, 7), (4, 8), (4, 9), (5, 8), \\
    &(5, 9), (6, 9), (10, 14), (10, 15), (11, 15), (12, 16), \\
    &(12, 17), (13, 16), (13, 17), (14, 17), (18, 22), (18, 23), \\
    &(19, 23), (20, 24), (20, 25), (21, 24), (21, 25), (22, 25), \\
    &(26, 31), (26, 33), (27, 30), (27, 32), (28, 31), (28, 33), \\
    &(29, 32), (29, 33), (2, 6), (2, 14), (2, 22), (2, 26)] \\
    G33p1 = [&(1, 6), (1, 7), (3, 7), (4, 8), (4, 9), (5, 8), \\
    &(5, 9), (6, 9), (10, 15), (10, 17), (11, 14), (11, 16), \\
    &(12, 15), (12, 17), (13, 16), (13, 17), (18, 23), (18, 25), \\
    &(19, 22), (19, 24), (20, 23), (20, 25), (21, 24), (21, 25), \\
    &(26, 31), (26, 33), (27, 30), (27, 32), (28, 31), (28, 33), \\
    &(29, 32), (29, 33), (2, 6), (2, 10), (2, 18), (2, 26)] \\
    G33p2 = [&(1, 6), (1, 7), (2, 6), (3, 7), (4, 8), (4, 9), \\
    &(6, 9), (10, 15), (10, 17), (11, 14), (11, 16), (12, 15), \\
    &(12, 17), (13, 16), (13, 17), (18, 23), (18, 25), (19, 22), \\
    &(19, 24), (20, 23), (20, 25), (21, 24), (21, 25), (26, 31), \\
    &(26, 33), (27, 30), (27, 32), (28, 31), (28, 33), (29, 32), \\
    &(29, 33), (5, 8), (5, 9), (5, 10), (5, 18), (5, 26)]
\end{align*}

\end{document}